\documentclass[a4paper,11pt]{article}
\usepackage{pos}
\usepackage{physics}
\usepackage{dsfont}
\usepackage{multirow}
\graphicspath{./}

\title{Baryonic screening masses in high temperature QCD}
\author*[a,b]{Pietro Rescigno}
\author[a,b]{Leonardo Giusti}
\author[c]{Tim Harris}
\author[a,b]{Davide Laudicina}
\author[b]{Michele Pepe}

\affiliation[a]{University of Milano-Bicocca,\\
  Piazza della Scienza 3, Milan, I-20126, Italy}

\affiliation[b]{INFN Milano-Bicocca,\\
  Piazza della Scienza 3, Milan, I-20126, Italy}
\affiliation[c]{Institute for Theoretical Physics, ETH Zürich,\\
Wolfgang-Pauli-Str. 27, 8093 Zürich, Switzerland}

\emailAdd{p.rescigno1@campus.unimib.it}
\emailAdd{leonardo.giusti@mib.infn.it}
\emailAdd{harrist@phys.ethz.ch}
\emailAdd{davide.laudicina@mib.infn.it}
\emailAdd{michele.pepe@mib.infn.it}

\abstract{We compute the screening masses of fields with nucleon quantum numbers for a wide range of temperatures between $T \sim 1$ GeV and $T\sim 160$ GeV. 
The computation has been performed by means of Monte Carlo simulations of lattice QCD with $N_f=3$ flavors of $O(a)$-improved Wilson fermions: we exploit a 
novel strategy which has recently allowed to determine for the first time non-singlet mesonic screening masses up to extremely high temperatures. The baryonic 
screening masses are measured with a few per-mille precision in the continuum limit, and percent deviations from the free theory result $3\pi T$ are clearly 
visible even at the highest temperatures. The observed degeneracy of the positive and negative parity state's screening mass, expected from Ward identities 
associated to non-singlet axial transformations, provides further evidence for the restoration of chiral symmetry in the high temperature regime of QCD.}

\FullConference{The 40th International Symposium on Lattice Field Theory (Lattice 2023)\\
July 31st - August 4th, 2023\\
Fermi National Accelerator Laboratory\\}

\begin{document}
\maketitle

\noindent\textbf{\large{1. Introduction }} \hspace{1mm} The study of Quantum ChromoDynamics (QCD) in the high temperature regime is a subject of particular 
interest for a variety of research fields,
 which range - to name a few areas - from the evolution of the early universe to the various experiments on heavy ion collisions such as SPS or RHIC.
From a theoretical point of view, the perturbative expansion of non-Abelian Gauge Theories is known to suffer from a severe infrared problem at 
high temperatures \cite{IR}, and it is thus necessary to employ non perturbative methods to investigate the properties of the Quark Gluon Plasma (QGP). 
Lattice simulations of QCD at finite temperature provide a solid way to detail the theory under these conditions from first principles, 
and to test the validity of the perturbative expansion. A class of observables that yield interesting information about the QGP are the 
screening masses in the various symmetry channels of the theory; these have been extensively studied on the lattice in the mesonic sector for 
temperatures around the critical temperature $T_c$ and up to $\sim 1$  GeV \cite{smdetar, smbazavov}, but they were only recently determined 
for very large temperature values up to $\sim 160$ GeV \cite{mesonpaper}. 
Conversely, lattice determinations of baryonic screening masses have, up to now, been restricted to low temperatures, and no continuum extrapolation 
has been performed, neither for the quenched case \cite{quenchbsm} nor in full QCD \cite{qcdbsm}. In these proceedings we present preliminary 
results from \cite{baryon}, with the aim to apply the strategy described in \cite{mesonpaper} for simulating QCD at very high 
temperatures to compute the baryonic screening masses in the high temperature range $1 \text { GeV} \lesssim T \lesssim 160 \text{ GeV}$.\\[2mm]
\noindent\textbf{\large{2. Definitions }} \hspace{1mm} Focusing on the baryonic sector of the theory, we consider an interpolating operator that carries the quantum numbers of a nucleon
\begin{equation}
	N(x) =\epsilon^{abc} \left(u^T_a(x) C \gamma_5 d_b(x)\right) d_c(x)\label{Nfield},
\end{equation}
and we define two correlation functions measured along the third spatial direction and projected to the lowest fermionic Matsubara frequency $\pi T$
\begin{equation}
	C_{N^\pm}( x_3)=\int dx_0 dx_1 dx_2 e^{i\pi \frac{x_0}{L_0}}\left\langle \Tr\left[ P_\pm N(x)\overline{N}(0) \right] \right\rangle\label{corr1}.
\end{equation}
The phase factor imposes the lowest fermionic Matsubara frequency, while also compensating the minus sign one gets when crossing the time boundary due 
to the antiperiodicity of the nucleon interpolating field. The subscripts $\pm$ refer to the positive $(N^+)$ and negative $(N^-)$ $x_3$ - parity nucleon states respectively, and the projectors on such states 
are given by $P_\pm=\left(\mathds{1} \pm \gamma_3 \right)/2$. The screening masses are defined as the coefficients that characterize the exponential decay 
of the correlation functions in eq. \eqref{corr1} for large $x_3-$ separations
\begin{equation}
	C_{N^\pm}(x_3) \underset{x_3\rightarrow\infty}{\sim}\exp{-m_{N^\pm}x_3}\left(1+O(e^{-\Delta_\pm x_3}) \right), \label{asymptotic}
\end{equation}
where $\Delta_\pm$ denotes the difference between the screening mass of the first excited state with nucleon quantum numbers and $m_{N^\pm}$. More 
precisely, the screening masses are defined by
\begin{equation}
	m_{N^\pm} \equiv \lim_{x_3\rightarrow\infty}-\frac{d}{dx_3}\ln{C_{N^\pm}(x_3)}\label{smdef}.
\end{equation}
A few remarks can be made at this point. On the one hand, in the non-interacting case - which by asymptotic freedom corresponds to the infinite temperature 
limit - the lowest energy contribution in eq. \eqref{corr1} comes from a state in which the three quarks all carry the lowest Matsubara frequency
 $\pi T$, therefore we expect
\begin{equation}
	\lim_{T\rightarrow \infty} m_{N^\pm} = 3\pi T. \label{infT}
\end{equation}
On the other hand, it can be shown that the opposite parity correlators in eq. \eqref{corr1} are related by a Ward Identity for SU(2)$_A$ transformations, that holds in the continuum theory, via
\begin{equation}
	C_{N^+}(x_3)\underset{\text{SU(2)}_A}{\longrightarrow} -C_{N^-}(x_3),\label{WI}
\end{equation}
which implies that if chiral symmetry is restored at high temperatures we expect a degeneracy of the positive and negative parity nucleon screening mass.\\[2mm]
\noindent\textbf{\large{3. Lattice setup }} \hspace{1mm} To measure the baryonic screening masses defined in eq. \eqref{smdef} we simulated QCD with $N_f=3$ 
flavors of $O(a)$-improved Wilson fermions in the chiral limit, 
regularized on a lattice of extension $L_0 \cross L^3$. The screening masses were computed for 12 temperature values labeled $T_0,\dots T_{11}$ ranging 
from around $\sim 1$ GeV up to $\sim 165$ GeV, see table \ref{tab1} for the precise values and other simulation details. 
The values of the bare gauge coupling, the critical values of the bare quark masses and the values of the Sheikoleslami-Wohlert improvement coefficient 
were taken from works by the ALPHA collaboration \cite{alpha1, alpha2, alpha3, alpha4} (we refer to \cite{mesonpaper} for a detailed description of the 
strategy used to simulate QCD at such high temperatures). For each temperature, we have simulated up to four lattices of different spacing $a$, 
corresponding to $L_0/a=4,6,8,10$, while the spatial extent of the lattice in each direction was always set to $L/a=288$. With an aspect 
ratio $L/L_0$ ranging from $\sim 20$ up to $\sim50$, finite volume effects are expected to be safely negligible; we are also performing dedicated 
runs on smaller spatial volumes to explicitly check that the values of the screening masses are compatible for different volumes. In our setup 
the thermal theory is defined in a moving frame by imposing shifted boundary conditions \cite{gm1,gm2,gm3} on the quark and gluon fields.\\[2mm]

\begin{figure}[h!]
	\centering
	\includegraphics[scale=0.45]{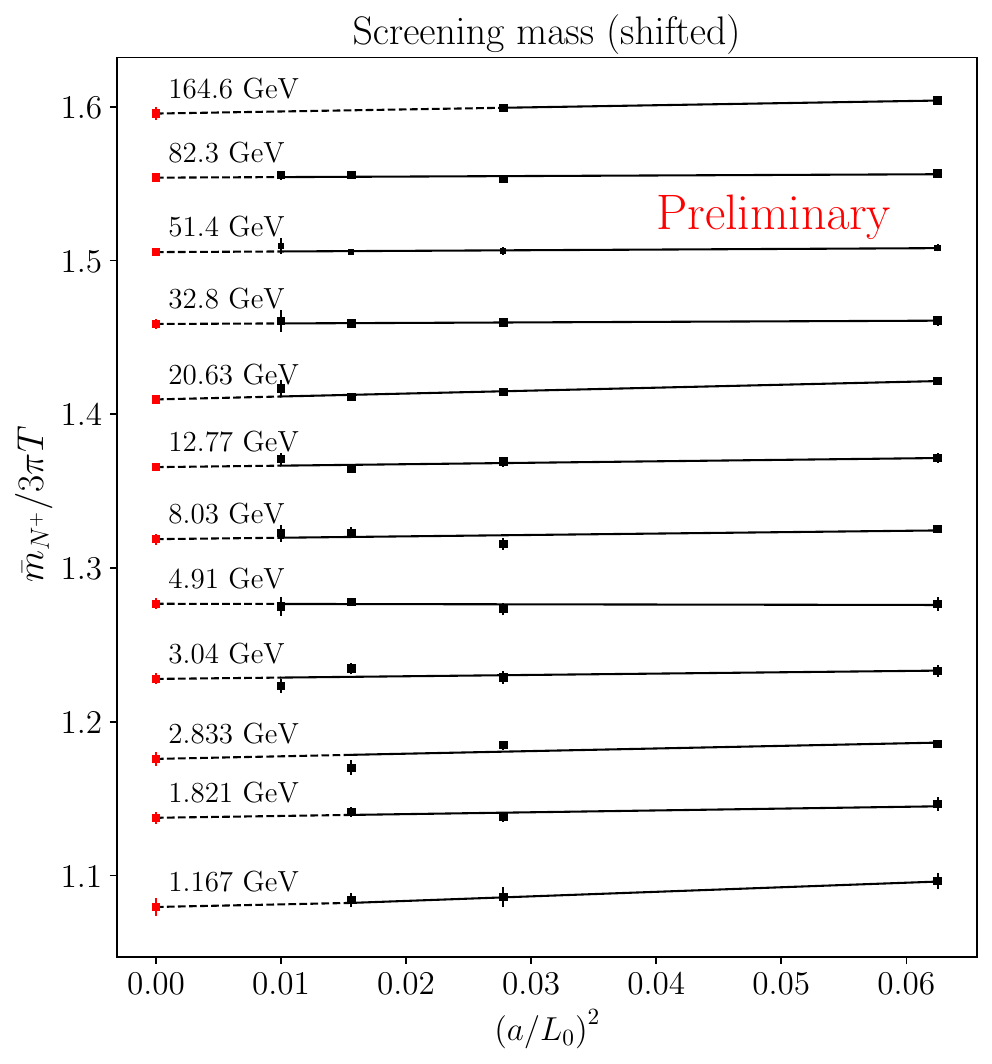}\hfill\includegraphics[scale=0.45]{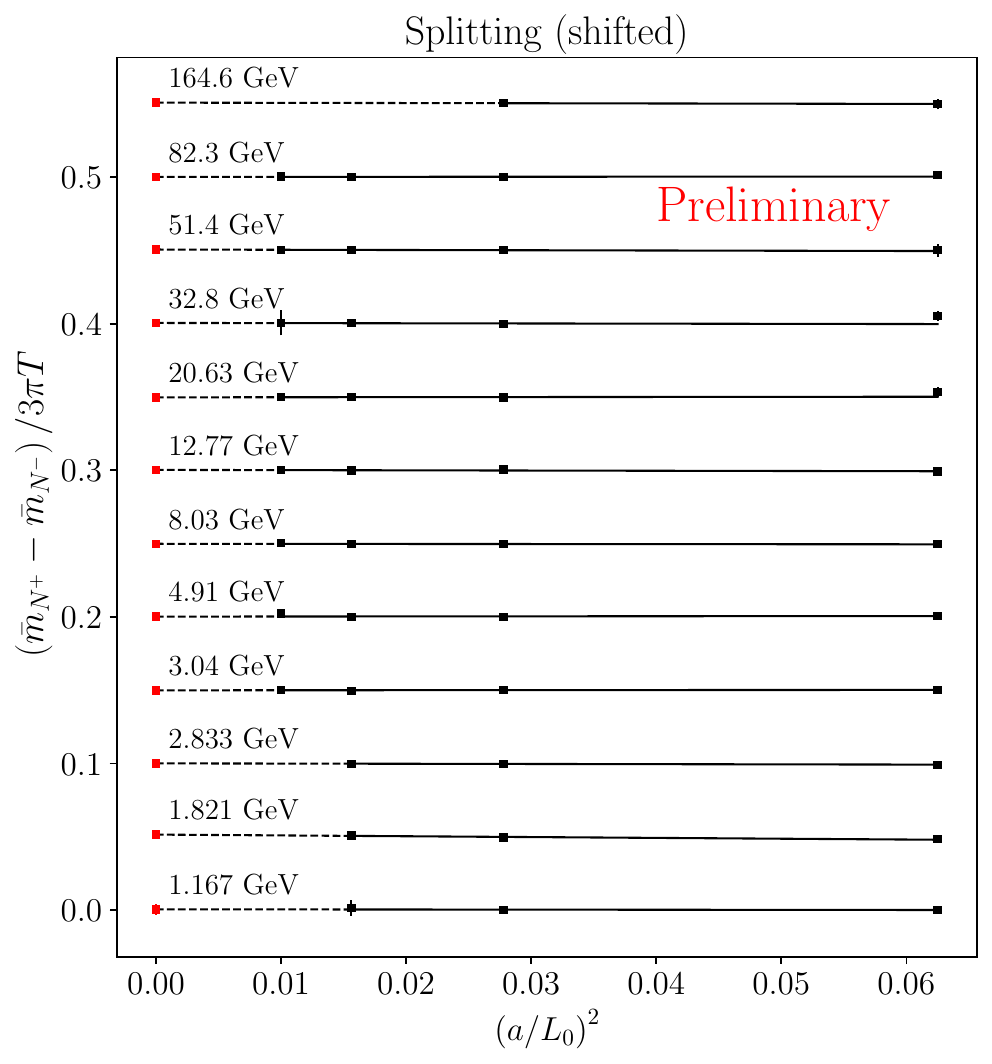}
	\caption{Continuum extrapolation of the positive parity screening mass (left) and the parity splitting (right), divided by the free theory value 
	$3\pi T$, for all temperatures. Data for $T_i$ is shifted upwards by $0.05(11-i)$ for better readability.\label{contfig}}
\end{figure}

\begin{table}[h!]
	    \centering            
	    \scriptsize
            \begin{tabular}{|c c | c c|}
            \hline
            &  &  &       \\[-0.125cm]
            $T$ & $T$[GeV] & $L_0/a$ & $n_{\rm mdu}$ \\[-0.125cm]
            &  &  &     \\
            \hline\hline  
            \multirow{4}{*} {\footnotesize $T_0$} & \multirow{4}{*} {\footnotesize 164.6(5.6)}          
            & \multirow{4}{*}{\shortstack[c]{4\\6}} & \multirow{4}{*}{\shortstack[c]{300\\390}}\\
	    & & & \\
	    & & & \\
	    & & & \\
            \hline                                        
            \multirow{4}{*} {\footnotesize $T_1$} & \multirow{4}{*} {\footnotesize 82.3(2.8)}
            &  4 & 300    \\
            & &  6 & 310   \\
            & &  8 & 500   \\
            & & 10 & 500  \\
            \hline                                        
            \multirow{4}{*} {\footnotesize $T_2$} & \multirow{4}{*} {\footnotesize 51.4(1.7)}
            &  4 & 300 \\
            & &  6 & 320 \\
            & &  8 & 490  \\
            & & 10 & 500 \\
            \hline                                        
            \multirow{4}{*} {\footnotesize $T_3$} & \multirow{4}{*} {\footnotesize 32.8(1.0)}
            &  4 & 300 \\
            & &  6 & 340 \\
            & &  8 & 490 \\
            & & 10 & 500 \\
            \hline                                        
	    \end{tabular}
	    \hspace{2mm}
	    \begin{tabular}{|c c | c c|}
	    \hline
            &  &  &       \\[-0.125cm]
            $T$ & $T$[GeV] & $L_0/a$ & $n_{\rm mdu}$ \\[-0.125cm]
            &  &  &     \\
            \hline\hline 
            \multirow{4}{*} {\footnotesize $T_4$} & \multirow{4}{*} {\footnotesize 20.63(63)}
            &  4 & 440   \\
            & & 6 & 310  \\
            & & 8 & 490 \\
            & & 10 & 500  \\
            \hline                                        
            \multirow{4}{*} {\footnotesize $T_5$} & \multirow{4}{*} {\footnotesize 12.77(37)}
            &  4 & 310   \\
            & & 6 & 310  \\
            & & 8 & 500  \\
            & & 10 & 500  \\
            \hline
	    \multirow{4}{*} {\footnotesize $T_6$} & \multirow{4}{*} {\footnotesize 8.03(22)}
            &  4 & 300  \\
            & & 6 & 320  \\
            & & 8 & 500  \\
            & & 10 & 500  \\
            \hline                                        
            \multirow{4}{*} {\footnotesize $T_7$} & \multirow{4}{*} {\footnotesize 4.91(13)}
            &  4 & 320  \\
            & & 6 & 310  \\
            & & 8 & 500  \\
            & & 10 & 500  \\
            \hline 

	    \end{tabular}
	    \hspace{2mm}
	    \begin{tabular}{|c c | c c|}                               
	    \hline
            &  &  &       \\[-0.125cm]
            $T$ & $T$[GeV] & $L_0/a$ & $n_{\rm mdu}$ \\[-0.125cm]
            &  &  &       \\
            \hline\hline           
                                                   
	    \multirow{4}{*} {\footnotesize $T_8$} & \multirow{4}{*} {\footnotesize 3.040(78)}
            &  4 & 320   \\
            & & 6 & 300   \\
            & & 8 & 500  \\
            & & 10 & 500 \\
            \hline
            \multirow{4}{*} {\footnotesize $T_9$} & \multirow{4}{*} {\footnotesize 2.833(68)}          
            &  \multirow{4}{*}{\shortstack[c]{4\\6\\8}} & \multirow{4}{*}{\shortstack[c]{400\\390\\390}} \\
            & &  &  \\
            & &  &  \\
	    & & &  \\
            \hline                                        
            \multirow{4}{*} {\footnotesize $T_{10}$} & \multirow{4}{*} {\footnotesize 1.821(39)}
           &  \multirow{4}{*}{\shortstack[c]{4\\6\\8}} & \multirow{4}{*}{\shortstack[c]{410\\400\\400}} \\
            & &   &    \\
            & &   &    \\
           & & &  \\
            \hline                                        
            \multirow{4}{*} {\footnotesize $T_{11}$} & \multirow{4}{*} {\footnotesize 1.167(23)}
            &  \multirow{4}{*}{\shortstack[c]{4\\6\\8}} & \multirow{4}{*}{\shortstack[c]{400\\390\\400}} \\
            & &  &   \\
            & &  &   \\
           & & &  \\ 
	   \hline
           \end{tabular}
           \caption{Features of the various simulations. The temperature labels and their values in GeV are reported, as well as the 
	Euclidean time interval in lattice units and the number $n_{\rm mdu}$ of MDU's generated (after thermalization of the Markov chain). The correlators were always computed on 4 local sources per configurations, except for the $L_0/a=4,6$ lattices at 
		$T_8$ (8 sources) and the $L_0/a=10$ lattice at $T_8$ (5 sources). A number $n_{\rm skip}=10$ of MDU's was always skipped between consecutive measurements.
		\label{tab1}}
\end{table}
\noindent\textbf{\large{4. Correlator and Noise-to-Signal Ratio}} \hspace{1mm}
\begin{figure}[h!]
	\centering
	\includegraphics[scale=0.6]{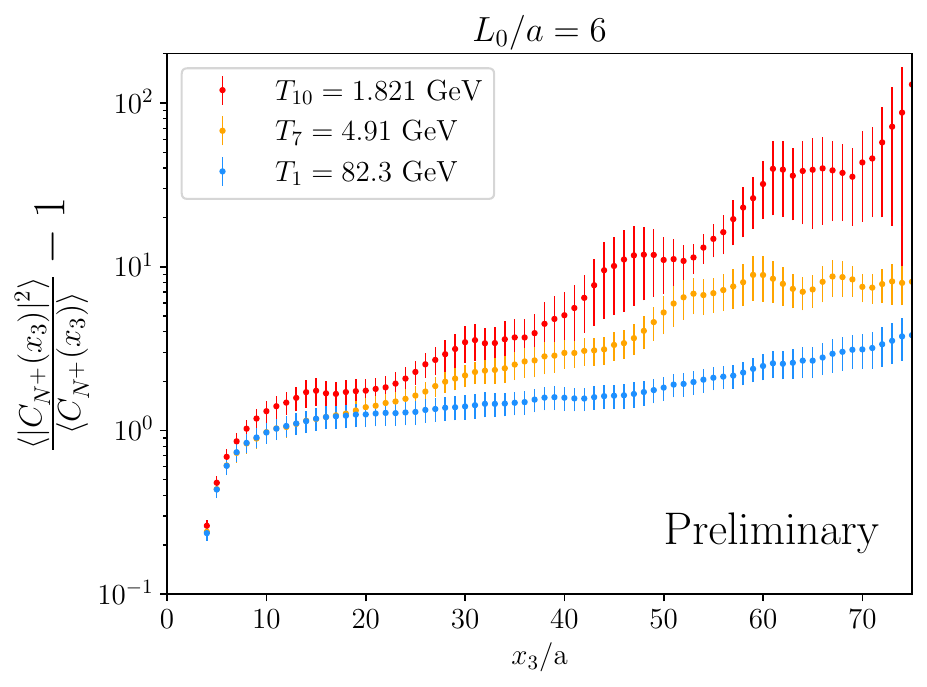}
	\caption{Ratio of the variance of the correlator to its expectation value squared (positive parity) as a function of the separation $x_3$ for three 
	different temperatures. The lattice geometry was $6\cross(288)^3$ in all cases. The higher the temperature, the slower the NSR grows with distance.\label{NSRplot}}
\end{figure}
Screening masses are obtained from the long-distance behavior of the correlator via eq. \eqref{smdef}. Hadronic correlators obtained from simulations of lattice QCD 
notoriously suffer from an exponential Noise-to-Signal Ratio (NSR) problem \cite{parisi84, lepage89}, i.e. the relative statistical error on the correlator grows 
exponentially with the source separation. While this feature severely hinders 
the determination of hadron masses at zero temperature, one can argue that at high temperatures a baryonic correlator measured along a 
spatial direction like the one in eq. \eqref{corr1} does not suffer from an exponential problem. Indeed, for large separations the ratio of the variance of 
$C_{N^\pm}(x_3)$ to its expectation value squared goes like
\begin{equation}
	\frac{\langle |C_{N^\pm}(x_3)|^2\rangle - \langle C_{N^\pm}(x_3)\rangle^2}{\langle C_{N^\pm}(x_3)\rangle^2}\sim \exp{(2m_{N^\pm}-3m_{P})x_3}, \label{NSR}
\end{equation}
where $m_P$ denotes the pseudoscalar mesonic screening mass. As eq. \eqref{infT} states, in the infinite temperature limit $T\rightarrow +\infty$ the 
baryonic screening masses approach the value $3\pi T$, and similarly the value of the mesonic screening mass tends to $2\pi T$; the coefficient in the exponent of 
eq. \eqref{NSR} thus tends to zero in this limit and the relative error on the correlator remains constant as $x_3$ increases. Since we are considering high but finite 
temperatures, we still observe an exponential increase of the NSR with distance, but at a much smaller rate than what one observes at zero temperature, and we are 
thus able to maintain a satisfactory statistical accuracy in the large distance region we are interested in. Figure \ref{NSRplot} displays the l.h.s. of eq. 
\eqref{NSR} for positive parity as a function of $x_3$, as obtained from our simulations, for three different temperature values on a $6\cross (288)^3$ lattice.\\[2mm]
\noindent\textbf{\large{5. Screening mass estimation }}\hspace{1mm}
\begin{figure}[h!]
	\centering
	\includegraphics[scale=0.6]{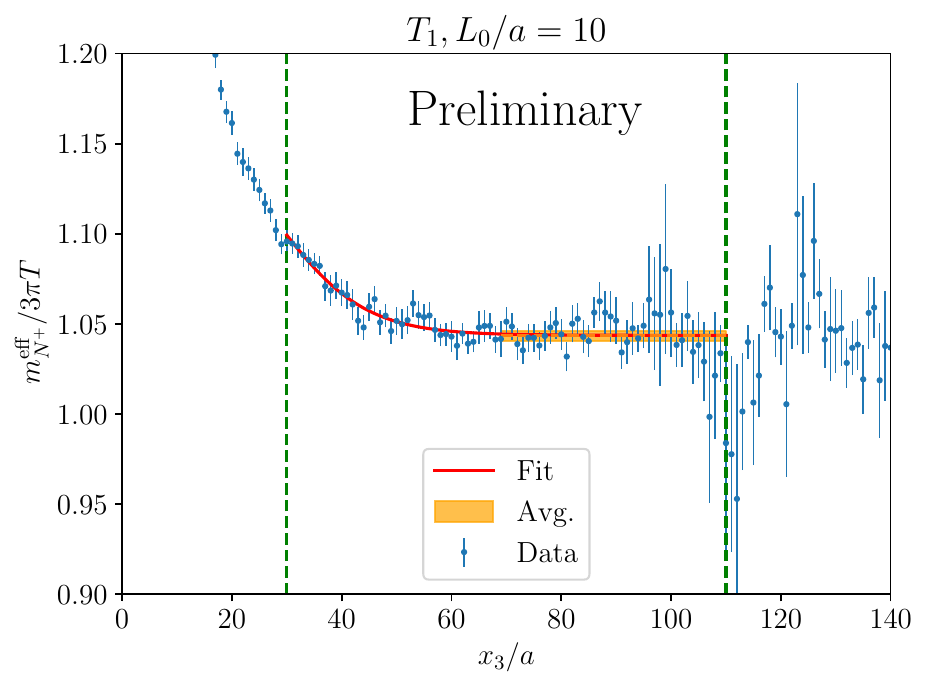}
	\caption{Effective screening mass values (positive parity) for a $10\cross(288)^3$ lattice at $T_1$. Dashed vertical lines indicate the fit interval, the red solid line 
		is the fit function and the orange band is the weighted average and its error in the interval it was computed.\label{effmassplot}}
\end{figure} 
From the correlator values we define an effective screening mass $am_{N^\pm}^{\rm eff}(x_3)\equiv - \ln \left[C_{N^\pm}(x_3+a)/C_{N^\pm}(x_3)\right]$, 
which for large distances plateaus around the value of the screening mass up to statistical fluctuations. We fit the effective mass values to a constant plus a correction 
due to the presence of an additional excited state in the spectral decomposition of $C_{N^\pm}(x_3)$, and we adjust the fit window to have a good fit quality as well as 
a non-negligible contribution from the excited state. From the fit results we determine the point after which the contribution of the excited state 
becomes negligible within our statistical precision, and we estimate the screening mass by taking a weighted average from this point up to when we 
eventually lose signal due to the exponential problem (for noisier datasets at the lower temperatures, we estimate the screening mass directly from the fit).
A representative effective mass dataset, together with the fit function and weighted average, is displayed in figure \ref{effmassplot}. 
To reduce discretization effects, we analytically computed the tree-level baryonic screening mass at finite lattice spacing 
$m^{\rm Tree}_{N^\pm}(L_0/a)$, and defined a tree-level improved screening mass as $\overline{m}_{N^\pm}(L_0/a)\equiv m_{N^\pm}(L_0/a) - 
\left(m_{N^\pm}^{\rm Tree}(L_0/a)-3\pi T \right)$. Continuum limit extrapolations for the positive parity screening mass and the difference of 
positive and negative parity screening masses are summarized in figure \ref{contfig}. The flatness of the interpolations and the good quality of the
fits to a constant plus quadratic term in $(a/L_0)$ signal the effectivness of the tree-level improvement of our data and the $O(a)$-improvement at 
the level of the action respectively.
The values of the screening masses in the continuum (red points in left panel of figure \ref{contfig}) are obtained with a statistical precision of a 
few parts per-mille, and within this accuracy the positive and negative parity screening masses are in perfect agreement at all temperatures and lattice 
spacings, as well as in the continuum (right panel of figure \ref{contfig}). This is a clear manifestation of chiral symmetry restoration in high 
temperature QCD.\\[2mm]
\noindent\textbf{\large{6. Temperature dependence }}\hspace{1mm}
Finally, we investigated the temperature dependence of $m_{N^\pm}/3\pi T$ by studying it as a function of $\hat{g}^2(T)=\left(\frac{9}{8\pi^2} 
\ln\frac{2\pi T}{\Lambda_{\overline{ \rm MS}}} +\frac{4}{9\pi^2} \ln\left( 2\ln \frac{2\pi T}{\Lambda_{\overline{ \rm MS}}}\right)\right)^{-1}$. For our 
purposes, we could have chosen any function of the temperature, but to facilitate the comparison with potential future perturbative predictions, we chose 
the 2 loop coupling constant in the $\overline{\rm MS}$ scheme renormalized at a scale $2 \pi T$. We have interpolated our data for $m_{N^\pm}/3\pi T$ 
with a polynomial in $\hat{g}$, and to reproduce the expected free theory result of eq. \eqref{infT} we set the constant term equal to 1. We observe that 
including a single additional $O(\hat{g}^2)$ term in the fit ansatz does not yield a satisfactory description of the curvature in our data, whereas 
including an additional higher power term $O(\hat{g}^3)$ or $O(\hat{g}^4)$ provides a much better quality fit. A cubic interpolation, together with our 
continuum extrapolated data for the positive parity screening mass, is displayed in figure \ref{finalplot}. As a consistency check, if we leave the constant 
term free during the fit and we include both a $O(\hat{g}^2)$ and a higher power term (either $O(\hat{g}^3)$ or $O(\hat{g}^4)$), we obtain a value for the 
constant term which is well compatible with 1.
\begin{figure}[h!]
	\centering
	\includegraphics[scale=0.6]{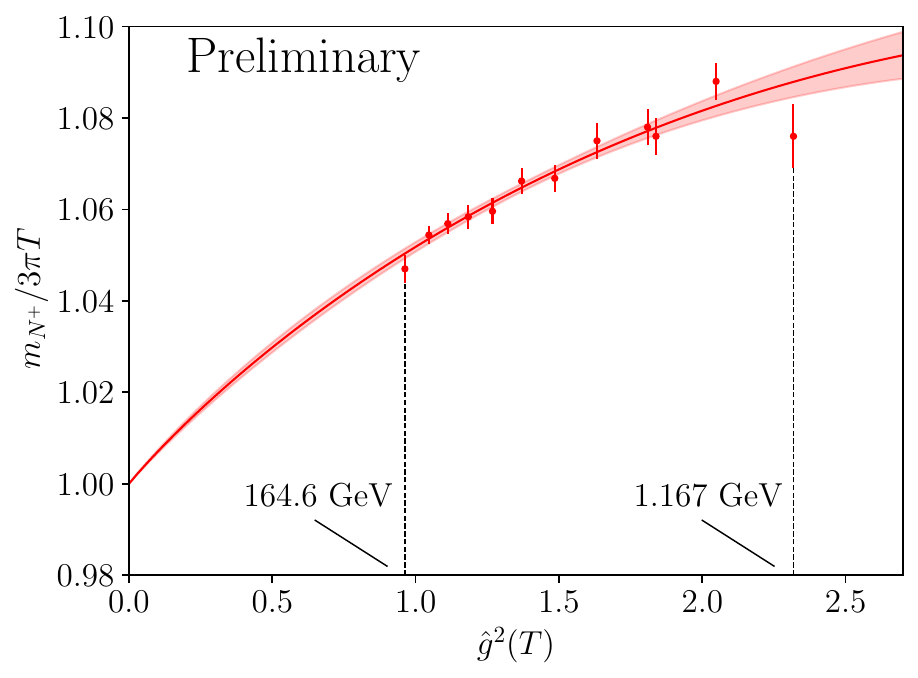}
	\caption{Continuum values for the positive parity screening mass (red points) as a function of the temperature dependent coupling defined in section 6. 
	The solid red line is an interpolation of the data to a cubic polynomial. \label{finalplot}}
\end{figure}\\[2mm]
\noindent\textbf{\large{7. Conclusions }}\hspace{1mm} We have presented the first continuum extrapolated results for baryonic screening masses in the 
high temperature range $1$ GeV $\lesssim T\lesssim 160$ GeV. We were able to demonstrate that at these temperatures the exponential problem for baryonic 
screening correlators is much milder than at zero temperature, which makes a computation of this kind feasible. The screening mass is obtained with a 
precision of a few parts per mille in the continuum, which makes it possible to appreciate deviations from the free theory result $3\pi T$ ranging from 
$4\%$ to $8\%$. The degeneracy of the positive and negative parity screening mass within our accuracy clearly signals the 
restoration of chiral symmetry in the high temperature regime of QCD. Our analysis of the temperature dependence of the results seems to suggest that only 
including an $O(\hat{g}^2)$ correction to the leading order free-theory result $3\pi T$ is not enough to reproduce the behaviour of the screening mass in 
the whole temperature range.\\[2mm]
\noindent\textbf{\large{Aknowledgements }}\hspace{1mm} We wish to thank Mikko Laine for several discussions on the topic of this paper. We acknowledge PRACE 
for awarding us access to the HPC system MareNostrum4 at the Barcelona Supercomputing Center (Proposals n. 2018194651 and 2021240051) and EuroHPC for the access 
to the HPC system Vega (Proposal n. EHPC-REG-2022R02-233) where most of the numerical results presented in this paper have been obtained. We also thank CINECA 
for providing us with computer- time on Marconi (CINECA- INFN, CINECA-Bicocca agreements). The R\&D has been carried out on the PC clusters Wilson and Knuth 
at Milano-Bicocca. We thank all these institutions for the technical support. This work is (partially) supported by ICSC – Centro Nazionale di Ricerca in 
High Performance Computing, Big Data and Quantum Computing, funded by European Union – NextGenerationEU.
%%%%%%%%%%%%%%%%%%%%%%%%%%%%%%%%%%%%%%%%%%%%%%%%%%%%%%%%%%%%%% REFERENCES %%%%%%%%%%%%%%%%%%%%%%%%%%%%%%%%%%%%%%%%%%%%%%%%%%%%%%%%%%%%%%%%%%%%%%%%%%%%%%%%%

\end{document}